\documentclass[11pt]{elsarticle}

\usepackage[colorlinks=true,
            linkcolor=green,
            urlcolor=green,
            citecolor=blue]{hyperref}
\usepackage[usenames,dvipsnames]{color}
\usepackage{makeidx}
\usepackage[none]{hyphenat}
\usepackage{graphicx,epsfig}
\usepackage{amsmath,amssymb}
\usepackage{fancyhdr,lastpage} 
\usepackage{xcolor}
\journal{Solid State Communications}
\begin{document}
\begin{frontmatter}
\title{Origin of the mid-infrared peaks in the optical conductivity of LaMnO$_{3}$}
\author[a,b]{Purevdorj Munkhbaatar}
\author[a,c]{Kim Myung-Whun\corref{cor1}}
\ead{ mwkim@jbnu.ac.kr}
\cortext[cor1]{To whom correspondences should be addressed}
\address[a]{Institute of Photonics and Information Technology, Jeonbuk National University, 54896 Republic of Korea}
\address[b]{Department of Physics, National University of Mongolia, Ulaanbaatar, 210646 Mongolia}
\address[c]{Department of Physics, Jeonbuk National University, 54896 Republic of Korea}
\date{\today}
\begin{abstract}
We calculated the infrared conductivity spectrum of orbitally ordered LaMnO$_3$ in phonon frequency and overtone frequency ranges. We considered orbital exchange, Jahn-Teller electron-phonon coupling, and phonon--phonon coupling. The fundamental excitation of the phonon-coupled orbiton was only Raman active, not infrared active, while its overtone modes were both Raman and infrared active. Our calculations reproduced the small peaks near 1300 cm$^{-1}$ observed both in Raman scattering and infrared conductivity spectra, as consistent with previous experimental results. 
\end{abstract}
\begin{keyword}
LaMnO$_3$ \sep infrared conductivity  \sep orbiton
\end{keyword}
\end{frontmatter}

In LaMnO$_3$, new low-energy elementary excitations can appear due to orbital ordering\cite{Khomskii,Cyrot,Ishihara}. Saitoh \textit{et al}. observed a weak signal in a Raman scattering experiment with LaMnO$_3$ and presented it as evidence of a new elementary excitation, or orbital wave\cite{Saitoh}. Shortly after this study,  weak peaks (mid-infrared peaks) were found in the same energy as the Raman peak in the optical (or infrared) conductivity spectrum ($\sigma$ ($\omega$)) \cite{Gruninger}. Since they were observed at the same energies, it can be thought that the mid-infrared peaks of $\sigma$($\omega$) may also be due to the fundamental mode of the orbiton. However, in a centrosymmetric system, vibrational modes should be either infrared active or Raman active, but not both. Owing to this contradiction of parity selection rules, the cause of the mid-infrared peaks in both Raman and $\sigma$($\omega$) spectra has remained the subject of exploration over the past 20 years.

The contradiction of the selection rules can be resolved by taking into account the  the complex combinations of two or more phonons. Gruninger \textit{et al}. interpreted the mid-infrared peaks of $\sigma$($\omega$) as the harmonic mode (two phonons) of low-energy phonons. However, as Saitoh \textit{et al}. pointed out, it was not clear which combinations of phonons can peak at the same energy in both Raman and $\sigma$($\omega$) spectra\cite{Saitoh2}. For the reason, it is not persuasive to assign the $\sigma$($\omega$) peak as a simple two-phonon peaks.

We pointed out that the 130, 140, and 160 meV peaks observed in the Raman scattering are not three independent orbital modes but rather the combined modes of orbital wave and phonon in our last study\cite{munkh}. Accepting our assignment, the mid-infrared peaks observed in $\sigma$($\omega$) can be understood also as the combined modes of orbital wave and phonon. The direct optical excitation of a typical orbital wave is not allowed, because the charge distribution in the excited state is symmetric with no net dipole moment. However the process of cooperative absorption of phonons and orbitons can be optically acceptable because phonon generation can induce a lower symmetry than the steady-state lattice\cite{Mizuno,Lorenzana}.

In this paper, we calculated the optical conductivity by taking into account
the harmonic excitation of the Raman-active, phonon-coupled orbiton. We
found that the infrared active modulation may occur in the orbital exchange
Hamiltonian because the exchange interaction is dependent on the displacement
of oxygen ions by the oscillating external electric field. Our calculation results
demonstrate that by this infrared active modulation, the overtones of Raman
active modes can appear as the optical conductivity peaks in the mid-infrared
region.

We calculated the energy dispersion curve using Brink's Hamiltonian\cite{van den Brink}: 
\begin{equation} \label{GrindEQ__1_} 
H=H_{orb}+H_{e-ph}+H_{ph}.   
\end{equation} 
Here, we assumed that the spin degrees of freedom were frozen. The first term corresponds to the super-exchange interaction between neighboring \textit{i}-th and \textit{j}-th orbitals and is expressed as follows:
\begin{equation} \label{GrindEQ__2_} 
H_{orb}=J\sum_{{\left\langle ij\right\rangle }_{\mathrm{\Gamma }}}{T^{\mathrm{\Gamma }}_iT^{\mathrm{\Gamma }}_j},   
\end{equation} 
where $J$ is the exchange coupling constant and $\mathrm{\Gamma }$ is the \textit{a} or \textit{b} axis. The orbital operator $T^{\mathrm{\Gamma }}_i$ is defined as $T^{\mathrm{a/b}}_i=({\tau }^{\mathrm{z}}_i\pm \sqrt{3}{\tau }^{\mathrm{x}}_i)/2$ using the pseudo-spin operator ($\tau $). The plus (minus) sign indicates the \textit{a}(\textit{b}) axis direction. 

\begin{figure}[t]
{\centering
\includegraphics[scale=0.65]{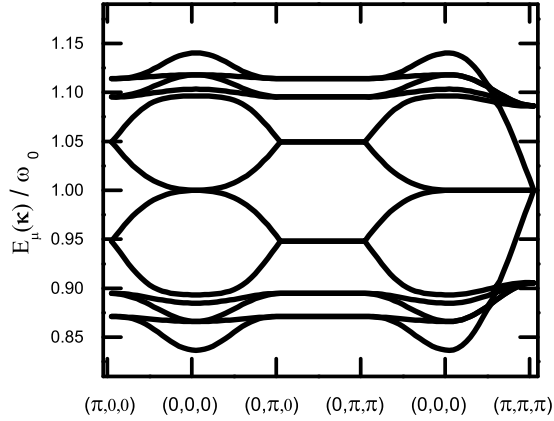} 
\par}
\caption{Dispersion curves of the phonon-coupled orbital wave.  The parameter values for obtaining the curves are chosen as $j=\omega_0$, $g/\omega_0=0.2$ and $\omega_1/\omega_0=-0.1$.}
\label{fig1}
\end{figure}

The second term corresponds to the Jahn-Teller type electron-phonon
coupling, expressed as follows:
\begin{equation} \label{GrindEQ__3_} 
H_{e-ph}=g\sum_i{\left({\tau }^{\mathrm{z}}_iQ_{3i}+{\tau }^{\mathrm{x}}_iQ_{2i}\right)},   
\end{equation} 
where $g$ is the electron-phonon coupling constant. $Q_2$ and $Q_3$ are phonon operators that describe two Jahn-Teller modes with e${}_{g}$ symmetry. The third term corresponds to the phonon contributions as 
\begin{equation} \label{GrindEQ__4_} 
H_{ph}={\omega }_0\sum_i{\left({a_{\mathrm{2i}}}^{\mathrm{\dagger }}a_{2i}+{a_{\mathrm{3i}}}^{\mathrm{\dagger }}a_{3i}\right)+{\omega }_1\sum_{{\left\langle ij\right\rangle }_{\mathrm{\Gamma }}}{Q^{\mathrm{\Gamma }}_iQ^{\mathrm{\Gamma }}_j}},   
\end{equation} 
where ${\omega }_0$ is the local Jahn-Teller phonon frequency for the $Q_{2}$ and $Q_{3}$ modes. We have assumed that the energies of the two phonons are the same to simplify the calculation. ${\omega }_1$is the nearest-neighbor coupling between the phonons, and $Q^{{\Gamma }}_i$ represents the coupled Jahn-Teller mode along a crystal axis $\mathrm{\Gamma }\mathrm{=}a$ or $b$. Figure~\ref{fig1} shows the energy dispersion derived from the Hamiltonian. More details of the calculation can be found in our previous study\cite{munkh}. 

The dispersion is similar to that of Saitoh \textit{et al}.\cite{Saitoh}. The difference is that multiple satellite dispersions are formed around the main orbital super-exchange dispersions; these satellite dispersions are created by phonon-phonon coupling\cite{munkh}. Because of the coupling, more possibilities are allowed for creating the mid-infrared excitations than the simple orbital exchange case.

\begin{figure}[t]
{\centering
\includegraphics[scale=0.2]{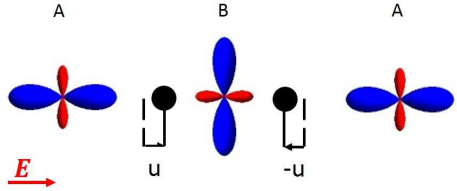} 
\par}
\caption{A schematic diagram representing the modulation of super-exchange coupling {${J}$} under the influence of external electric field $E$. The Jahn-Teller distortion cause oxygen (O) ion displacement {$u$}. A and B represent the orbital subgroup
of Mn-ions.}
\label{fig2}
\end{figure}

We calculated the infrared conductivity by using the Lorenzana-Sawatzky method\cite{Lorenzana}. First, we investigated the influence of the optical phonon in the super-exchange Hamiltonian by modifying super-exchange coupling $J$ as a function of electric field $E$ and oxygen displacement $u$ (See Appendix for the details.):

\begin{equation} \label{GrindEQ__5_} 
H_{orb}\mathrm{=}J(\boldsymbol{E},u)\sum_{\left\langle ij\right\rangle }{T_iT_j}.     
\end{equation} 
\begin{equation} \label{GrindEQ__6_} 
J\left(\boldsymbol{E},u\right)=J_0+q_IE\ u.      
\end{equation}
Here $J_0$ is the static super-exchange coupling constant and $q_I$ is effective charge. 

We reestablished the exchange Hamiltonian in the presence of the electric field by using $q_I$. The Hamiltonian determines how light, lattice, and orbitals interact, so we call it the interaction Hamiltonian $H_I$. We assumed that the electric field is polarized along the \textit{x}-axis (or \textit{a}-axis). The electric field is described by the vector potential $A_{\boldsymbol{x}}=A_0e^{i(\boldsymbol{k}\cdot\boldsymbol{r}\boldsymbol{-}\omega t)}$, assuming that the light propagates perpendicularly to the \textit{ab}-plane. In this case, the phase factor $e^{i\boldsymbol{k}\cdot\boldsymbol{r}}$ can be neglected. The interaction Hamiltonian along the \textit{a}-axis is as follows: 
\begin{equation} \label{GrindEQ__10_} 
H_I=q_I\ u\sum_{{\left\langle ij\right\rangle }_{\mathrm{a}}}{T^{\mathrm{a}}_iT^{\mathrm{a}}_j}\frac{\omega A_{\boldsymbol{x}}}{c} .           
\end{equation} 

To simplify the calculation, we assumed that the oxygen ion displaces along the Mn-O-Mn bond direction in the ideal lattice as the electric field perturbs the system. As Figure~\ref{fig2} shows, at any moment, the oxygen ion displacement can be in two directions. This means that the displacement of oxygen ions should be described by dividing them into two lattice groups, a$_{1}$ and a$_{2}$. We designated the group moving in the negative direction as a$_{1}$ and the group moving in the positive direction as a$_{2}$. According to this designation, the interaction Hamiltonian becomes:
\begin{equation} \label{GrindEQ__11_} 
H_I=q_I\ u\left(-\sum_{{\left\langle ij\right\rangle }_{\mathrm{a1}}}{T^{\mathrm{a}}_iT^{\mathrm{a}}_j}+\sum_{{\left\langle ij\right\rangle }_{\mathrm{a2}}}{T^{\mathrm{a}}_iT^{\mathrm{a}}_j}\right)\frac{\omega A_{\boldsymbol{x}}}{c}.    
\end{equation} 
With this Hamiltonian, we can create the current operator as: 
\begin{equation} \label{GrindEQ__12_} 
j(\omega )=\left(-\sum_{{\left\langle ij\right\rangle }_{\mathrm{a1}}}{T^{\mathrm{a}}_iT^{\mathrm{a}}_j}+\sum_{{\left\langle ij\right\rangle }_{\mathrm{a2}}}{T^{\mathrm{a}}_iT^{\mathrm{a}}_j}\right)q_I\ u\frac{\omega }{c}.         
\end{equation} 
We calculated the optical conductivity using the Kubo formula:
\begin{equation} \label{GrindEQ__13_} 
\sigma \left(\mathrm{\omega }\right)=Im\left(\frac{c^2}{V}\sum_e{\frac{\left\langle g\mathrel{\left|\vphantom{g j(\mathrm{\omega }) e}\right.\kern-\nulldelimiterspace}j(\mathrm{\omega })\mathrel{\left|\vphantom{g j(\mathrm{\omega }) e}\right.\kern-\nulldelimiterspace}e\right\rangle \left\langle e\mathrel{\left|\vphantom{e j(\mathrm{\omega }) g}\right.\kern-\nulldelimiterspace}j(\mathrm{\omega })\mathrel{\left|\vphantom{e j(\mathrm{\omega }) g}\right.\kern-\nulldelimiterspace}g\right\rangle }{\mathrm{\omega }\ (E_e-E_g-\hslash \mathrm{\omega }+i\ \mathrm{\gamma }\mathrm{)}}}\right).   
\end{equation} 
where $E_{g(e)}$ is the energy of the ground (excited) state, $V$ is the volume of the model system, and ${c}$ is the speed of light. $\mathrm{\gamma }$ is a phenomenological spectral broadening factor, and we introduce dipole operator $d=q_I u(-\sum_{{\left\langle ij\right\rangle }_{a_1}}{T^{\mathrm{a}}_iT^{\mathrm{a}}_j}+\sum_{{\left\langle ij\right\rangle }_{a_2}}{T^{\mathrm{a}}_iT^{\mathrm{a}}_j})$. Optical conductivity can be expressed by the dipole operator:
\begin{equation} \label{GrindEQ__14_} 
\sigma \left(\omega \right)=Im\left(\frac{1}{V\hslash }\sum_e{\frac{\omega \left\langle g\mathrel{\left|\vphantom{g d e}\right.\kern-\nulldelimiterspace}d\mathrel{\left|\vphantom{g d e}\right.\kern-\nulldelimiterspace}e\right\rangle \left\langle e\mathrel{\left|\vphantom{e d g}\right.\kern-\nulldelimiterspace}d\mathrel{\left|\vphantom{e d g}\right.\kern-\nulldelimiterspace}g\right\rangle }{({\omega }_{eg}-\omega +i\ \mathrm{\gamma }\mathrm{)}}}\right),    
\end{equation} 
where ${\omega }_{eg}\equiv {(E}_e-E_g)/\hslash $. 

$d$ becomes $\sum_{{\left\langle ij\right\rangle }_{\mathrm{a}}}q_I u{{\xi\left(\theta \right)}_{ij}\left(a_{1i}+{a_{\mathrm{1i}}}^{\mathrm{\dagger }}\right)\left(a_{1j}+{a_{\mathrm{1j}}}^{\mathrm{\dagger }}\right)+O(a^3)}$ with ${ \xi\left(\theta \right)}_{ij}$=$\pm (1+2\ \mathrm{cos}(2\theta ))/16$. Here $a_{1i}$ is the Holstein boson operators\cite{munkh}. The orbital angle $\theta $ determines the shape of the orbital and the strength of the exchange coupling. The plus (minus) sign corresponds to the lattice a$_{1}$ (a$_{2}$). The dipole operator has no first-order terms of the Holstein bosons, which means that the fundamental modes of infrared conductivity are not allowed, i.e. ${\sigma}^{(1)} \left(\mathrm{\omega }\right)=0.$ The second-order terms of the Holstein bosons correspond to the two-particle absorption of the overtone modes. The second-order terms give the overtone infrared conductivity $\sigma^{(2)}$:
\begin{equation} \label{GrindEQ__15_} 
{\sigma }^{(2)}\left(\mathrm{\omega }\right)=\sum_e{Im({(\mathrm{\omega}-{\omega }_{eg}+i\ \mathrm{\gamma}\mathrm{)}}^{-1})\frac{{q_I}^2u^2}{V\hbar }}{\left|\left\langle e\mathrel{\left|\vphantom{e \sum_{{\left\langle ij\right\rangle }_{\mathrm{a}}} {\color{red} \xi} {\left(\theta \right)_{ij}\left(a_{1i}+{a_{\mathrm{1i}}}^{\mathrm{\dagger }}\right)\left(a_{1j}+{a_{\mathrm{1j}}}^{\mathrm{\dagger }}\right)} g}\right.\kern-\nulldelimiterspace}\sum_{{\left\langle ij\right\rangle }_{\mathrm{a}}}{{ \xi\left(\theta \right)}_{ij}\left(a_{1i}+{a_{\mathrm{1i}}}^{\mathrm{\dagger }}\right)\left(a_{1j}+{a_{\mathrm{1j}}}^{\mathrm{\dagger }}\right)}\mathrel{\left|\vphantom{e \sum_{{\left\langle ij\right\rangle }_{\mathrm{a}}}{{J\left(\theta \right)}_{ij}\left(a_{1i}+{a_{\mathrm{1i}}}^{\mathrm{\dagger }}\right)\left(a_{1j}+{a_{\mathrm{1j}}}^{\mathrm{\dagger }}\right)} g}\right.\kern-\nulldelimiterspace}g\right\rangle \right|}^2.     
\end{equation} 

We used the Bogoliubov transformation to calculate the matrix elements \cite{munkh}. The result is as follows:
\begin{eqnarray} 
\nonumber\sigma^{(2)}(\omega) &=&\sum_{\boldsymbol{k}\mu \mu'}\mbox{Im}\left[\left(\omega -\{E_{\mu}(\boldsymbol{k})+E_{\mu'}(-\boldsymbol{k})\}/\hbar +i\gamma\right)^{-1}\right]\frac{{q_I}^2u^2}{V\hbar}\left|\sum_{\nu \nu'}{\rho }_{\nu\nu'}(k_x)\left(V_{\nu \mu }(\boldsymbol{k})\right.\right.\\
\label{GrindEQ__16_} && \left. \left.+W_{\nu \mu }^*(\boldsymbol{k})\right)\left( V_{\nu '\mu '}(-\boldsymbol{k})+{W_{\nu '\mu '}}^*(-\boldsymbol{k})\right)\right|^2. 
\end{eqnarray} 
Here ${\rho }_{\nu \nu '}\equiv \pm (1+2\ \mathrm{cos}(2\theta ))(1-\mathrm{cos}(k_x))/16$. $V_{\nu \mu }(\boldsymbol{k})$ and $W_{\nu \mu }(\boldsymbol{k})$ are Bogoliubov transformation coefficients connecting the boson operator for the \textit{$\nu$-}th orbiton in a unit cell to that of the $\mu$-th eigenmode with energy $E_{\mu }\left(\boldsymbol{k}\right)$. 

\begin{figure*}
{\centering
\includegraphics[scale=0.7]{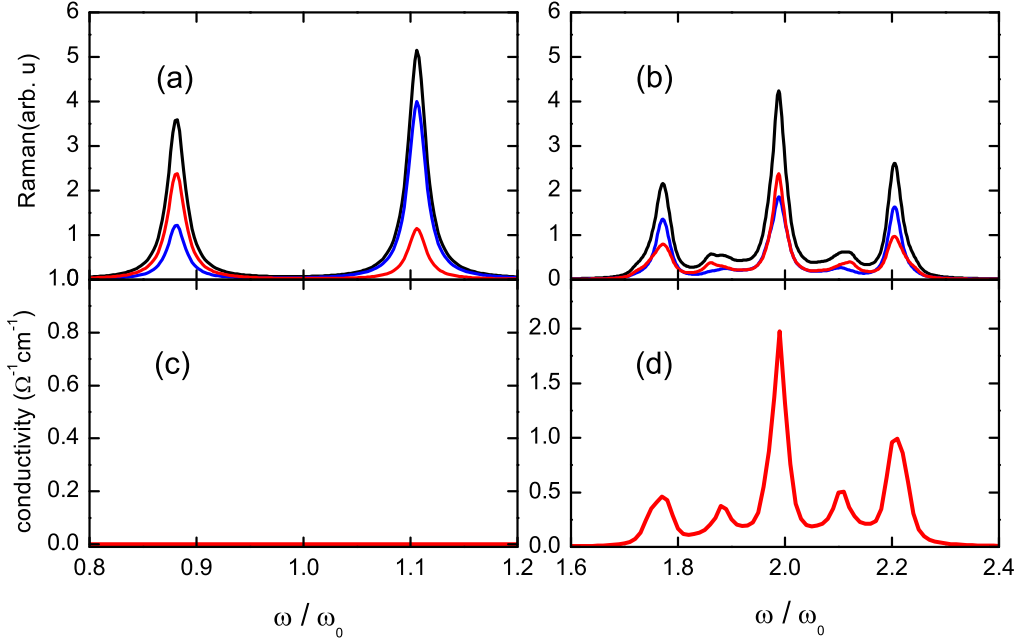} 
\par}
\caption{(a,b) Raman scattering spectrum: (a) fundamental modes and (b) first overtone modes. (c,d) Infrared conductivity spectrum: ({\color{red} c}) fundamental modes and (d) the first overtone modes.}
\label{fig3}
\end{figure*}

Figure~\ref{fig3} shows the numerical calculation results of Raman scattering and infrared conductivity $\sigma \left(\mathrm{\omega }\right)$. We used the results of our previous study for the Raman scattering spectrum. We chose 40$\times $40 lattice cells for the numerical calculations of $\sigma \left(\mathrm{\omega }\right)$. The parameters for the calculation were ${\omega }_2\mathrm{=}{\omega }_0$,$\mathrm{\ }{\omega }_1=-0.1\,{\omega }_0$, $J=0.5\,{\omega }_0$, $g=0.2\,{\omega }_0$, and $\theta =\pi /2$, the same as our previous study\cite{munkh}. The lattice parameters were $a=5.80\,\AA$, $ b=5.61\,\AA$, $ c=7.79\,\AA$. The system volume was $V=Nabc=40\times 40\times 5.8\times 5.6\times 7.8\approx 4.0\times {10}^{-25}{m}^{3}$, and the dipole was $q_Iu=-0.2\times 1.6\times {10}^{-19}{C}\times {0.2}\AA=6.4\times {10}^{-31}\mbox{C m}.$ The broadening factor was $\gamma = 0.01\, \omega_0$.

Before discussion, we need to review the literature. The orbital wave proposed by Saitoh \textit{et al}. is a Raman active excitation which requires switching between pseudospins without parity change. This transition cannot directly contribute to $\sigma$($\omega$) due to parity selection rules, so the orbital wave cannot be directly related to the small peak in the mid-infrared $\sigma$($\omega$). Saitoh \textit{et al}. explained that the mid-infrared peaks in $\sigma$($\omega$) are a spin-allowed d-d electronic excitation that becomes infrared active through the disruption of local inversion symmetry due to impurities or defects\cite{Saitoh}. In LaMnO$_3$ though, the energy scale of the d-d electron excitation is known to occur above 1 eV, so such an assignment is not convincing\cite{mwkim}.

An alternative interpretation is that the peaks in Raman and $\sigma$($\omega$) are due to multi-phonons. Mid-infrared peaks of $\sigma$($\omega$) appear between 130 meV and 160 meV \cite{Gruninger}; this corresponds to approximately twice the phonon energy between 60 meV and 80 meV \cite{Saitoh2,Fedorov,Paolone,Tobe}. Such a coincidence between energies leads us to the idea that the mid-infrared peaks are the overtones of the fundamental phonons. However, a simple doubling of the infrared active modes does not exactly match the frequencies of the mid-infrared peaks. It is also not easy to find the appropriate combinations of two infrared-active modes that produce the mid-infrared peaks. 

In our previous study, we proposed the phonon-coupled orbiton, a Bogoliubov boson. We found that the orbiton becomes infrared active like the Jahn-Teller phonon involves the orbiton excitation.  The Jahn-Teller distortion changes the parity of the orbiton. Such a process has not been considered in any other previous studies. In our model, the energy of the orbiton is twice smaller than that of Saitoh \textit{et al}., and the bare orbiton is strongly coupled with the Jahn-Teller phonon\cite{munkh}. The phonon-coupled orbiton is different from the 'bare orbiton' proposed by Saitoh \textit{et al}., but we will refer to 'phonon-coupled orbiton' as 'orbiton' for convenience in the following.

The small Raman scattering peaks were explained in our previous work in terms of the overtones of the phonon-coupled orbiton. Figure~\ref{fig3} shows the calculated Raman scattering and $\sigma(\omega)$ spectrum for low energy (near 70 meV) and overtone energy (near 140 meV) regions. Figures~\ref{fig3}(a) and ~\ref{fig3}(b) are the Raman scattering spectra presented in our previous paper\cite{munkh}. In the figure, only ($x,x$) polarization results in which the Mn-O-Mn bond and the electric field direction are parallel are shown for comparison with $\sigma$($\omega$). The rest of the polarization results are not much different and can be found in previous studies\cite{munkh}.

Figure~\ref{fig3}(a) corresponds to the fundamental Raman modes. ${\omega}_0$ is the bare orbiton energy, with two peaks appearing on both sides. We guessed that ${\omega}_0$ is about 70 meV. Both peaks are strongly coupled with the $Q_{2}$ and $Q_{3}$ phonons and interact with each other. The peak at the lower frequency is similar to the classical phonon since a higher proportion of the lattice vibration contributes to the Bogoliubov transformation coefficient of the peak. On the other hand, the peak at the higher frequency is also coupled to the Jahn-
Teller phonon, but it is reasonable to call it an orbiton because of the higher
proportion of the orbital contribution in the Bogoliubov transformation.

Figure~\ref{fig3}(b) shows the first Raman overtone modes of Figure~\ref{fig3}(a). Three peaks are Raman active, and their frequencies correspond to almost twice the fundamental mode frequencies. The lowest frequency peak is due to two-phonon excitation, and the highest energy is due to two-orbiton excitation. The peak in the middle is due to the phonon-assisted orbiton. Our calculations reproduced the results of the Raman scattering experiment\cite{Saitoh}.

We also calculated the infrared conductivity spectrum, $\sigma$($\omega$).  Figure~\ref{fig3}(c) shows no peaks, which is consistent with the fact that the fundamental orbiton mode cannot be observed in $\sigma$($\omega$). On the other hand, Figure~\ref{fig3}(d) shows peaks similar to the Raman active modes at almost the same energy. These are the two-phonon excitation, phonon-assisted orbiton excitation, and two-orbiton excitation from low energy, as in the Raman scattering spectrum. Also, the strengths of the mid-infrared peaks in $\sigma$($\omega$) are much weaker ($<$ 1\%) than those of the main phonons, consistent with the experiments\cite{Gruninger}. 

The calculation results show that the combination of Raman fundamental modes can produce mid-infrared peaks in the infrared conductivity spectrum as well as Raman overtone modes simultaneously. If the Raman fundamental modes are simple classical phonons, any combination of the Raman fundamental modes could not produce the mid-infrared peaks in the infrared conductivity spectrum because of the symmetry selection. The overtone modes confirm that the Raman fundamental modes are not simple classical phonons but orbtial wave excitation (phonon-coupled orbitons) in which electronic degrees of freedom participate.

The light-matter interaction mechanisms producing Raman overtones and infrared overtones need not be the same. The Raman overtone peaks can be due to both virtual d-d excitation and virtual p-d excitation. On the other hand, virtual d-d excitation can not activate the overtone peaks in the infrared conductivity spectrum by the linear electric field because the electric dipole can not be formed by the excitation. Instead virtual p-d excitation can make the infrared overtone modes active in the linear infrared conductivity because the virtual p-d excitation can form the effective electric dipole (See Appendix for the details). Even if the overtone generation mechanisms of Raman and infrared are different, the two spectra are calculated based on the same energy dispersion curve. Although the microscopic light-matter interactions are different, the resonant energy of two processes (Raman overtone and infrared overtone) must be the same because the ground state and final state of the processes are the same. The energy coincidence of the overtone peaks demonstrates that the origins of the weak mid-infrared peaks are the same in Raman scattering and the infrared conductivity.

In addition, our two-orbiton scenario may show the hint to explain the large
temperature-dependent energy shift of the mid-infrared Raman peaks. The
large energy shift has been suggested as one of the reasons why the midinfrared
Raman peaks cannot be attributed to simple multi-phonon scattering\cite{Saitoh}. In contrast to the weak dependence of the low-energy phonon mode energies,
the mid-infrared peaks show substantial temperature-dependent energy
shifts. However, we think this cannot be concrete evidence supporting the
assignment of Saitoh \textit{et al}. A single orbiton is usually generated near the
gamma point, whereas a two-orbiton can combine various crystal momentums
on the dispersion curve. For this reason, among the posible excitations
participating in the two-orbiton process, some modes may be excited from/to
the band edge. The dispersion curve can change more significantly near the
band edge than near the gamma point under the influence of the orbital/spin
ordering. This is possibly a cause of the larger temperature-dependent energy
shift of the two-orbiton.

In conclusion, we found that the overtone modes of the fundamental
phonon-coupled orbiton can be both Raman active and infrared active at
the mid-infrared energy. We think that the fundamental modes are in the vicinity of 611 cm$^{-1}$, and that the modes represent the coupled excitation of $Q_{2}$- and $Q_{3}$-phonons and orbital excitation. The three (or more) peaks observed in mid-infrared $\sigma$($\omega$) as well as Raman scattering are combinations of the fundamental modes, such as two-phonon, one-phonon plus one-orbiton, and two-orbiton.

\section*{Acknowledgement}

This research was supported by the basic research program of the National Research Foundation of Korea (Grants No. 2019R1F1A1061994). PM was supported by the Korea research fellowship program through the National Research Foundation of Korea (Grants No. 2019H1D3A1A01071216).

\appendix
\section{superexchange modulation}

We used the Lorenzana-Sawatzky (LS) method for calculating the infrared conductivity. Though the LS method was derived for the magnetic interaction of cuprates, it can be also applied to the pseudo-spin interaction in LaMnO$_3$ with a little modification of the exchange constant and the effective charge. Followings show the details of our modification.

To take account the effect of the electric field on the superexchange interaction, we considered two possible systems for Hubbard models: (1) Mn-Mn and (2) Mn-O-Mn. In the followings we estimate the modulation induced in the superexchange interaction by the external electric field.

(1) \textbf{Mn-Mn} \\
In LaMnO$_3$ there are inter-site virtual transitions (d$^4_i$ d$^4_j\longrightarrow $d$^3_i$ d$^5_j$) contributing to the orbital superexchange \cite{klaliullin}. To drive the effective exchange Hamiltonian, we considered a spineless four-band (Mn-Mn) effective Hubbard model in the external electric field ($\mathcal{E}$):
\begin{eqnarray}
\nonumber H&=&t_{dd}(d_{1\Uparrow}^{\dagger}d_{2\Uparrow}+d_{2\Uparrow}^{\dagger}d_{1\Uparrow})+U(d_{1\Uparrow}^{\dagger}d_{1\Uparrow}d_{1\Downarrow}^{\dagger}d_{1\Downarrow}+d_{2\Uparrow}^{\dagger}d_{2\Uparrow} d_{2\Downarrow}^{\dagger}d_{2\Downarrow})\\
\nonumber & &+(\varepsilon_d+er_1	\mathcal{E})(d_{1\Uparrow}^{\dagger}d_{1\Uparrow}+d_{1\Downarrow}^{\dagger}d_{1\Downarrow})+(\varepsilon_d+er_2\mathcal{E})(d_{2\Uparrow}^{\dagger}d_{2\Uparrow}+d_{2\Downarrow}^{\dagger}d_{2\Downarrow}).
\end{eqnarray}
Here $d^{\dagger}_{i,\sigma}$ creates an electron with $\sigma$ orbital at $i$th Mn-site. $\Uparrow$ ($\Downarrow$) corresponds $3x^2-r^2$($y^2-z^2$) orbital in a Mn ion, $e$ is the electron charge, and $t_{dd}$ is the inter-site hopping integral between two Mn ions. $\varepsilon_d$ is the band energy of the Mn d-orbital and $U$ is repulsive energy of doubly occupied in a Mn ion. We assumed that the band energies are modulated linearly by the electric dipole interaction as the position of Mn ion ($r_i$) changes. 

Matrix form of the Hamiltonian can be decomposed into four blocks and basis states as following: 
\begin{equation*}
(2 \varepsilon_d+e\mathcal{E}(r_1+r_2))\quad\mbox{and}\quad d_{1\Uparrow}^{\dagger} d_{2\Uparrow}^{\dagger}\vert 0\rangle,
\end{equation*}

\begin{equation*}
(2 \varepsilon_d+e\mathcal{E}(r_1+r_2))\quad\mbox{and}\quad d_{1\Downarrow}^{\dagger} d_{2\Downarrow}^{\dagger}|0\rangle,
\end{equation*}

\begin{equation*}
\left(
\begin{array}{cc}
2 \varepsilon_d+e\mathcal{E}(r_1+r_2) & t_{dd}  \\
t_{dd} & 2 \varepsilon_d+U+e\mathcal{E}(2r_2) 
\end{array}\right)\quad\mbox{and}\quad\left(
\begin{array}{c}
d_{1\Uparrow}^{\dagger} d_{2\Downarrow}^{\dagger}|0\rangle\\
d_{2\Uparrow}^{\dagger}d_{2\Downarrow}^{\dagger}|0\rangle
\end{array}\right),
\end{equation*}

\begin{equation*}
\left(
\begin{array}{cc}
2 \varepsilon_d+e\mathcal{E}(r_1+r_2) & t_{dd}  \\
t_{dd} & 2 \varepsilon_d+U+e\mathcal{E}(2r_1)  \\
\end{array}\right)\quad\mbox{and}\quad\left(
\begin{array}{c}
d_{1\Downarrow}^{\dagger}d_{2\Uparrow}^{\dagger}|0\rangle\\
d_{1\Downarrow}^{\dagger} d_{1\Uparrow}^{\dagger}|0\rangle
\end{array}\right),
\end{equation*}
where $|0\rangle$ is vacuum state.

We established the projection operator $P$ on a Mn ion single electron occupied subspace of a Hilbert space and $Q\equiv 1-P$ \cite{hubbard}. In the subspace, the Schr\"{o}dinger equation and the effective Hamiltonian are written by 
\begin{equation}\label{schr}
\widehat{H}(E)|\psi\rangle=E|\psi\rangle,
\end{equation}
where
\begin{equation}\label{effham}
\widehat{H}(E)=PH(1+(E-QH)^{-1}QH)P.
\end{equation}
Here $E=2 \varepsilon_d+e\mathcal{E}(r_1+r_2)$. The solutions of the effective Hamiltonian are following:
\begin{eqnarray*}
\vert \psi_1\rangle&=&d_{1\Uparrow}^{\dagger} d_{2\Uparrow}^{\dagger}\vert 0\rangle \Rightarrow E_1=2 \varepsilon_d+e\mathcal{E}(r_1+r_2),\\
\vert \psi_2\rangle&=&d_{1\Downarrow}^{\dagger} d_{2\Downarrow}^{\dagger}\vert 0\rangle \Rightarrow E_2=2 \varepsilon_d+e\mathcal{E}(r_1+r_2),\\
\vert \psi_3\rangle&=&d_{1\Uparrow}^{\dagger} d_{2\Downarrow}^{\dagger}\vert 0\rangle \Rightarrow E_3=E_1+\dfrac{t_{dd}^2}{U+e\mathcal{E}(r_2-r_1)},\\
\vert \psi_4\rangle&=&d_{1\Downarrow}^{\dagger} d_{2\Uparrow}^{\dagger}\vert 0\rangle \Rightarrow E_4=E_1+\dfrac{t_{dd}^2}{U+e\mathcal{E}(r_1-r_2)}.
\end{eqnarray*}
We can define the exchange energy as the energy difference between ferro-orbital state and antiferro-orbital state. We estimated the exchange energy constant between orbitals as $J \equiv E_1+E_2-E_3-E_4$. 

If we expand the $J$ to the second order in $\mathcal{E}$, then 
\begin{equation*}
    J\approx 2t_{dd}^2/U+e^2\,(r_1-r_2)^2\,\mathcal{E}^2t_{dd}^2/U^3.
\end{equation*}
In the absence of external filed, this orbital exchange is consistent with the result in Ref \cite{klaliullin} when the spins are aligned ferromagnetic. In the presence of external field, on the other hand, this superexchange can be only affected by the quadratic field. Hence the Mn-Mn Hubbard model can not contribute to the linear infrared conductivity.

(2) \textbf{Mn-O-Mn} \\
In LaMnO$_3$, the inter-site virtual transitions can be also achieved by involving the intermediate oxygen site. To drive the effective exchange Hamiltonian of the case, we considered a spineless five-band (Mn-O-Mn) effective Hubbard model in the external electric field ($\mathcal{E}$):
\begin{eqnarray}
\nonumber H&=&t_{pd}(d_{1\Uparrow}^{\dagger}p+p^{\dagger}d_{1\Uparrow}+d_{2\Uparrow}^{\dagger}p+p^{\dagger}d_{2\Uparrow})\\
\nonumber & &+(\varepsilon_d+er_1\mathcal{E})(d_{1\Uparrow}^{\dagger}d_{1\Uparrow}+d_{1\Downarrow}^{\dagger}d_{1\Downarrow})+(\varepsilon_d+er_2\mathcal{E})(d_{2\Uparrow}^{\dagger}d_{2\Uparrow}+d_{2\Downarrow}^{\dagger}d_{2\Downarrow})+(\varepsilon_p+er_p\mathcal{E}) p^{\dagger}p\\
\nonumber &&+U(d_{1\Uparrow}^{\dagger}d_{1\Uparrow} d_{1\Downarrow}^{\dagger}d_{1\Downarrow}+d_{2\Uparrow}^{\dagger}d_{2\Uparrow} d_{2\Downarrow}^{\dagger}d_{2\Downarrow}).
\end{eqnarray}
Here $d^{\dagger}_{i,\sigma}$ ($p^{\dagger}$) creates an electron with $\sigma$ orbital at $i$-site in Mn $d$ orbital (O $p_x$ orbital). $\Uparrow$ ($\Downarrow$) corresponds $3x^2-r^2$($y^2-z^2$) orbital in a Mn ion. $e$ is the electron charge. $t_{pd}$ is the inter-site hopping integral between an O ion and a Mn ion. $U$ is repulsive energy of doubly occupied Mn sites. We assumed that the band energies ($\varepsilon_d$ and $\varepsilon_p$) of Mn and O ion sites are modulated linearly by the electric dipole interaction as the position of ions ($r_i$) change.

Matrix form of the Hamiltonian has four blocks and basis states  as following: 

\begin{equation*}
E_0\quad\mbox{and}\quad d_{1\Uparrow}^{\dagger}p^{\dagger} d_{2\Uparrow}^{\dagger}\vert 0\rangle,
\end{equation*}

\begin{equation*}
\left(
\begin{array}{ccc}
E_0 & t_{pd} & t_{pd} \\
t_{pd} & 3 \varepsilon_d+U+e\mathcal{E}(2r_1+r_2) & 0 \\
t_{pd} & 0 & 3 \varepsilon_d+U+e\mathcal{E}(r_1+2r_2) 
\end{array}\right)\quad\mbox{and}\quad\left(
\begin{array}{c}
d_{1\Downarrow}^{\dagger}p^{\dagger} d_{2\Downarrow}^{\dagger}|0\rangle\\
d_{1\Downarrow}^{\dagger}d_{1\Uparrow}^{\dagger}d_{2\Downarrow}^{\dagger}|0\rangle\\
d_{1\Downarrow}^{\dagger} d_{2\Uparrow}^{\dagger}d_{2\Downarrow}^{\dagger}|0\rangle
\end{array}\right),
\end{equation*}

\begin{equation*}
\left(
\begin{array}{ccc}
E_0 & t_{pd} & 0 \\
t_{pd} & 3 \varepsilon_d+U+e\mathcal{E}(2r_2+r_1) & t_{pd} \\
0 & t_{pd} & 2 \varepsilon_d+\varepsilon_p+U+e\mathcal{E}(2r_2+r_p) 
\end{array}\right)\quad\mbox{and}\quad\left(
\begin{array}{c}
d_{1\Uparrow}^{\dagger}p^{\dagger} d_{2\Downarrow}^{\dagger}|0\rangle\\
d_{1\Uparrow}^{\dagger}d_{2\Uparrow}^{\dagger}d_{2\Downarrow}^{\dagger}|0\rangle\\
p^{\dagger}d_{2\Uparrow}^{\dagger}d_{2\Downarrow}^{\dagger}|0\rangle
\end{array}\right),
\end{equation*}

\begin{equation*}
\left(
\begin{array}{ccc}
E_0 & t_{pd} & 0 \\
t_{pd} & 3 \varepsilon_d+U+e\mathcal{E}(r_2+2r_1) & t_{pd} \\
0 & t_{pd} & 2 \varepsilon_d+\varepsilon_p+U+e\mathcal{E}(r_p+2r_1) 
\end{array}\right)\quad\mbox{and}\quad\left(
\begin{array}{c}
d_{1\Downarrow}^{\dagger}p^{\dagger} d_{2\Uparrow}^{\dagger}|0\rangle\\
d_{1\Downarrow}^{\dagger}d_{1\Uparrow}^{\dagger}d_{2\Uparrow}^{\dagger}|0\rangle\\
d_{1\Downarrow}^{\dagger} d_{1\Uparrow}^{\dagger}p^{\dagger}|0\rangle
\end{array}\right),
\end{equation*}
where $E_0\equiv 2 \varepsilon_d+\varepsilon_p+e\mathcal{E}(r_p+r_1+r_2)$ and $|0\rangle$ is the vacuum state. 

We established the projection operator method similar to the Mn-Mn case. In  the subspace of a Mn ion single electron occupied, the Schr\"{o}dinger equation and the effective Hamiltonian can be written by using eq.(\ref{schr})-(\ref{effham}). If we set $E=E_0$, then the solutions of the effective Hamiltonian are following:
\begin{eqnarray*}
\vert \psi_1\rangle&=&d_{1\Uparrow}^{\dagger}p^{\dagger} d_{2\Uparrow}^{\dagger}\vert 0\rangle \Rightarrow E_1=E_0,\\
\vert \psi_2\rangle&=&d_{1\Downarrow}^{\dagger}p^{\dagger} d_{2\Downarrow}^{\dagger}\vert 0\rangle \Rightarrow E_2=E_0+\frac{t_{pd}^2}{E_0-(3 \varepsilon_d+U+e\mathcal{E}(2 r_1+r_2))}+\frac{t_{pd}^2}{E_0-(3 \varepsilon_d+U+e\mathcal{E}(r_1+2r_2))},\\
\vert \psi_3\rangle&=&d_{1\Uparrow}^{\dagger}p^{\dagger} d_{2\Downarrow}^{\dagger}\vert 0\rangle \Rightarrow E_3=E_0+\dfrac{t_{pd}^2 (E_0- 2 \varepsilon_d-\varepsilon_p-U-e\mathcal{E}(2r_2+r_p) )}{(E_0-3 \varepsilon_d-U-e\mathcal{E}(2r_2+r_1))(E_0- 2 \varepsilon_d-\varepsilon_p-U-e\mathcal{E}(2r_2+r_p) )-t_{pd}^2},\\
\vert \psi_4\rangle&=&d_{1\Downarrow}^{\dagger}p^{\dagger} d_{2\Uparrow}^{\dagger}\vert 0\rangle \Rightarrow E_4=E_0+\dfrac{t_{pd}^2(E_0-3\varepsilon_d-\varepsilon_p-U-e\mathcal{E}(r_p+2r_1) )}{(E_0-3\varepsilon_d-U-e\mathcal{E}(r_2+2r_1))(E_0-3\varepsilon_d-\varepsilon_p-U-e\mathcal{E}(r_p+2r_1))-t_{pd}^2}.
\end{eqnarray*}
We definded the exchange constant between orbitals: $J \equiv E_1+E_2-E_3-E_4$. We expanded $J$ to the first order in $\mathcal{E}$: $$J\approx J_0+q_I\,u\,\mathcal{E}.$$  
where
\begin{equation*}\label{eq1} J_0 =\frac{2 t_{pd}^4}{(U+{\varepsilon}_{d}-\varepsilon_p) (-t_{pd}^2 + U(U+\varepsilon_d- \varepsilon_p))},
\end{equation*}
\begin{equation*}\label{eq2} 
q_I =\frac{ 2e t_{pd}^4 (-t_{pd}^2 + 2 U(U+\varepsilon_d-\varepsilon_p))}{(U+{\varepsilon}_{d}-{\varepsilon}_{p})^2 (-t_{pd}^2 + U(U+\varepsilon_d- \varepsilon_p))^2},
\end{equation*}
Here $u=(r_1+r_2)/2-r_p$ is the oxygen displacement with respect to the middle point of two Mn ions. We assumed that ${\varepsilon_d-\varepsilon_p }$ and $t_{pd}$ are negligible smaller than $U$, and $J_0\approx 0.02 U$: 
\begin{equation*} \label{GrindEQ__9_} 
q_I= \frac{ e J_0 (-t_{pd}^2 + 2 U(U+\varepsilon_d-\varepsilon_p))}{(U+{\varepsilon}_{d}-{\varepsilon}_{p}) (-t_{pd}^2 + U(U+\varepsilon_d- \varepsilon_p))}\approx 0.04e. 
\end{equation*}

In the absence of external field, again the orbital exchange becomes the same with the result of Ref \cite{klaliullin}. In the presence of the external electric field, however the superexchange can be linearly modulated by the field. We used the modulated superexchange to calculate the infrared conductivity.

\end{document}